\documentclass[12pt,eps]{article}

\usepackage{graphicx}
\usepackage{epstopdf}
\usepackage{amssymb}

\newcommand{\rmi}{\mathrm{i}}
\newcommand{\rmd}{\mathrm{d}}

\begin{document}
\thispagestyle{empty}

\begin{center}

{\Large\bf Relativistic Kinematics of Two-Parametric Riemann Surface in Genus Two}

\vspace{0.5cm}
A.V. Nazarenko\\
Bogolyubov Institute for Theoretical Physics,\\
14-b, Metrologichna Str., Kiev 03680, Ukraine\\
e-mail: nazarenko@bitp.kiev.ua

\vspace{0.3cm}
Yu.A. Kulinich\\
Astronomical Observatory, Ivan Franko National University of Lviv,\\
8, Kyryla and Methodia Str., Lviv 79005, Ukraine\\
e-mail: kul@astro.franko.lviv.ua

\begin{abstract}
It is considered a model of compact Riemann surface in genus two, represented
geometrically by two-parametric hyperbolic octagon with an order four automorphism and
described algebraically by the corresponding Fuchsian group. Introducing the
Fenchel--Nielsen variables, we compute the Weil--Petersson (WP) symplectic two-form for
parameter space and analyze the closed isoperimetric orbits of octagons. WP-Area in
parameter space and the canonical action--angle variables for the orbits are found.
Exploiting the ideas from the loop quantum gravity, we generate relativistic
kinematics by the Lorentz boost and quantize WP-area. We treat the evolution
in terms of global variables within the ``big bounce'' concept.

\vspace{0.2cm}
\end{abstract}

\end{center}


\newpage
\section{Introduction}

The Riemann surfaces are often used in the problems of string theory~\cite{DHP}, lower-dimensional
gravity~\cite{Lo,NR} and quantum geometry~\cite{Kash00}. Sometimes, chaotic behavior in nature
can be also related with non-trivial geometry \cite{Gutz,Naz}.

Here we study free geometrodynamics or kinematics of the surface in genus two associated with
hyperbolic octagon which is embedded into Poincar\'e disk and stable under rotation by $\pi/2$.
Identifying the opposite sides of such a domain, it is enough to use two real parameters in order to
describe the octagon geometry and the form of Fuchsian group generators. To generate a geometry
evolution, we operate by parameter space invariants instead of Riemannian metric which is unknown for us.
Note that the involution of similar surfaces and the associated generators were discussed in \cite{Sil,BS}.

Defining the Teichm\"uller space~\cite{IT} for a family of the given surfaces, we find the Fenchel--Nielsen
variables regarding as global coordinates on it and permitting us to endow the parameter space with the
Weil--Petersson (WP) symplectic two-form due to the Wolpert's theorem~\cite{Wo85}. Further, we demonstrate
that the parameter space can be densely covered by the closed curves of the constant octagon perimeter and
describe the diffeomorphism produced by isoperimetric constraint. We consider the set of isoperimetric
orbits as a tool for further parametrization and quantization~\cite{Hurt}.

We determine the canonically conjugate action--angle variables for isoperimetric orbits by identifying the
action with WP-area of domain bounded by the orbit in parameter space. This is a main point of our approach
using the integral characteristics in a contrast with the formalism based on local Fenchel--Nielsen
parameters of the surface \cite{Kash00}. In this way, we touch the problem of WP-area quantization, which
is similar to the one of the loop quantum gravity/cosmology~\cite{RS}.

In order to describe geometrodynamics and to perform area quantization, we extend the algebra of
action--angle variables up to generators of $\mathfrak{su}(1,1)$ associated with the Lorentz algebra in
(2+1)-dimensional space-time. Such an approach permits us to formulate the relativistic dynamics of Riemann
surface as a canonical transformation generated by the boost. It is realized in Section 4 and leads to
``big bounce'' scenario~\cite{APS,LM} in parameter space. We note that the quantization of the universe
represented by the Riemann surface in genus two was formulated algebraically in \cite{NR}.

\section{Model Octagon and Riemann Surface}

The Poincar\'e model of two-dimensional hyperbolic space is given by open disk
$\mathbb{D}=\left\{z=x+\rmi y||z|<1\right\}$ and the metric $\rmd s^2=4|\rmd z|^2/(1-|z|^2)^2$.
The geodesic in $(\mathbb{D},\rmd s^2)$ is an circle arc inside $\mathbb{D}$ with radius $R$ and
center at the point $z_0=\sqrt{1+R^2}\exp{(\rmi\phi)}$ lying beyond the unit disk.
In particular case, the geodesics emanating from the origin are the Euclidean straight
lines (diameters). All geodesics intersect the boundary $\partial\mathbb{D}$ orthogonally.

The group of all orientation-preserving isometries $\gamma$ of $(\mathbb{D},\rmd s^2)$,
denoted by ${\rm Isom^+(\mathbb{D})}$, acts via the M\"obius transformation:
\begin{equation}
z\mapsto \gamma[z]=\frac{uz+v}{\overline{v}z+\overline{u}},\quad
z\in\mathbb{D},
\end{equation}
where $u$ and $v$ satisfy relation $|u|^2-|v|^2=1$; $\bar u$, $\bar v$ are the complex
conjugates. Thus, it is convenient to identify a generator $\gamma$ with an element of group
\begin{equation}
{\rm SU(1,1)}=\left\{\left.
\left(\begin{array}{cc}
u&v\\
\overline{v}&\overline{u}
\end{array}\right)\right||u|^2-|v|^2=1
\right\}.
\end{equation}

The Riemann surface $S$ is understood here as a compact two-dimensional orientable manifold
with the metric of constant negative curvature. Such a surface is obtained from hyperbolic
simply connected octagon ${\cal F}$ in $\mathbb{D}$ via gluing opposite sides formed by
eight geodesic arcs, whose intersections serve as vertices.

\begin{figure}
\begin{center}
\includegraphics[width=4.2cm]{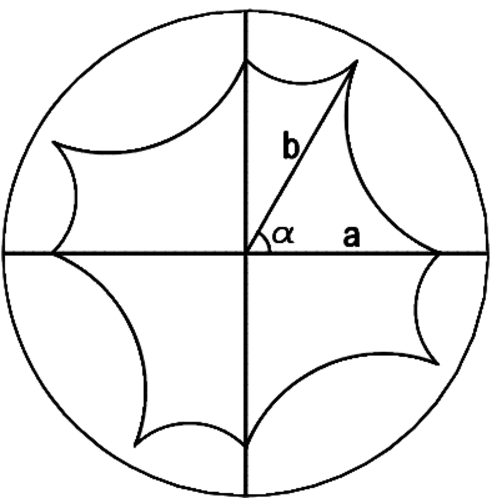}\qquad
\includegraphics[width=4.3cm]{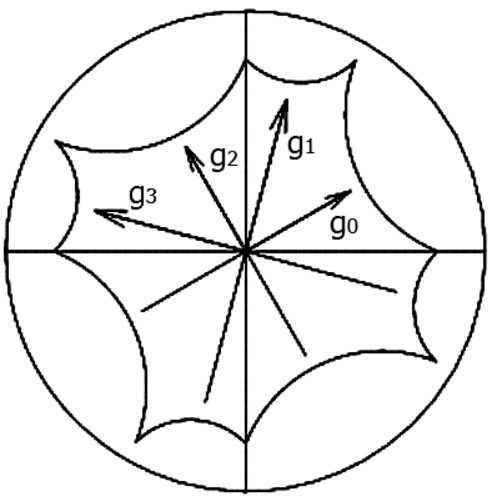}
\end{center}
\vspace*{-3mm}
\caption{\small Octagon with $a=0.8$, $\alpha=\pi/3$ and generators $g_k$ of
Fuchsian group.}
\end{figure}

Let the vertices be at the points $a\exp{(\rmi k\pi/2)}$, $b\exp{[\rmi(\alpha+k\pi/2)]}$
(Fig.~1, left panel), where $0<\alpha<\pi/2$, $0<a,b<1$ and $k=\overline{0,3}$. We also
require that the sum of inner angles of ${\cal F}$ equals to $2\pi$ and hyperbolic
${\rm Area}({\cal F})=4\pi$ in the case of the surface in genus two.

Choosing two parameters $(a,\alpha)$ as independent real variables, we find that
$b$ and the inner angle $\beta$ by vertices $a\exp{(\rmi k\pi/2)}$ are
\begin{equation}\label{b}
b=\left(\sqrt{2}a\cos{\tilde\alpha}\right)^{-1}, \quad
\tan{\beta}=\frac{1-a^2}{1-b^2}; \quad
\tilde\alpha=\alpha-\frac{\pi}{4}.
\end{equation}
These allow us to determine the region ${\cal A}$ of variety of parameters $(a,\alpha)$:
\begin{equation}
-\pi/4<\tilde\alpha<\pi/4,\quad
\left(\sqrt{2}\cos{\tilde\alpha}\right)^{-1}<a<1,
\end{equation}
which is shown in Fig.~2 below.

To obtain the regular hyperbolic octagon, we should put $a=2^{-1/4}$, $\alpha=\pi/4$.

In the case at hand, the octagon boundary $\partial{\cal F}$ is formed by geodesics of two
kinds (labeled by ``$\pm$'' below), which are completely determined by the radii $R_\pm$ and
the angles $\phi_\pm+k\pi/2$, $k=\overline{0,3}$, defining the positions of the circle (arc)
centers. Geometrical conditions result in parametrization:
\begin{equation}\label{R-phi}
R_\pm=\frac{1}{2a}\sqrt{T^2_\pm+(1-a^2)^2},\qquad
\phi_\pm=\arctan{\left[\left(\frac{T_\pm}{1+a^2}\right)^{\pm1}\right]},
\end{equation}
here $T_\pm=a^2\pm\tan{\tilde\alpha}$ and $0<\phi_+<\alpha<\phi_-<\pi/2$.

Here, we connect the model octagon ${\cal F}$ with the corresponding Riemann surface $S$ and
the Fuchsian group $\Gamma\subset{\rm Isom^+(\mathbb{D})}$ isomorphic to fundamental group $\pi_1(S)$.

Since the opposite sides of ${\cal F}$ have the same lengths, we can define isometry
$g_k\in{\rm Isom}^+(\mathbb{D})$ mapping geodesic boundary segment $s_{k+4}$ onto $s_k$
for all $k=\overline{0,3}$ (see Fig.~1, right panel). Identifying any $z\in s_{k+4}$ with
$g_k[z]\in s_k$, we obtain a closed surface. Four isometries $g_k$ and their inverses
$g^{-1}_k$ generate Fuchsian group $\Gamma$ with a single relation:
\begin{equation}
g_0 g^{-1}_1 g_2 g^{-1}_3 g^{-1}_0 g_1 g^{-1}_2 g_3={\rm id},
\end{equation}
and we then define surface $S$ as a quotient $\mathbb{D}/\Gamma$.

Our calculations give the dependence of $g_0$ and $g_1$ on $(a,\alpha)$:
\begin{eqnarray}
&g_0=N(a,\tilde\alpha)\left(
\begin{array}{cc}
	a(1-\tan{\tilde\alpha})& (a^2-\tan{\tilde\alpha})+\rmi(1-a^2)\label{g0}\\
	(a^2-\tan{\tilde\alpha})-\rmi(1-a^2) & a(1-\tan{\tilde\alpha})
\end{array}
\right),\\
&g_1=N(a,\tilde\alpha)\left(
\begin{array}{cc}
	a(1+\tan{\tilde\alpha})& (1-a^2)+\rmi(a^2+\tan{\tilde\alpha})\\
	(1-a^2)-\rmi(a^2+\tan{\tilde\alpha}) & a(1+\tan{\tilde\alpha})
\end{array}
\right),
\end{eqnarray}
here
$$
N(a,\tilde\alpha)=\frac{-\cos{\tilde\alpha}}{\sqrt{(1-a^2)(2a^2\cos^2{\tilde\alpha}-1)}}.
$$

The remaining generators are simply obtained by rotations:
\begin{equation}\label{gn}
g_{2,3}=R_\frac{\pi}{2}g_{0,1}R^{-1}_\frac{\pi}{2}, \quad
g^{-1}_k=R_{\pi}g_kR^{-1}_{\pi};
\end{equation}
$$
R_\varphi=\left(
\begin{array}{cc}
	\exp{\left(\rmi{\varphi}/{2}\right)}& 0\\
	0 & \exp{\left(-\rmi{\varphi}/{2}\right)}
\end{array}
\right).
$$

Let us mark a surface by generators of $\Gamma$. Two marked surfaces $(S,\Gamma)$ and
$(S^\prime,\Gamma^\prime)$ are called marking equivalent if there exists an isometry
$\gamma: S\to S^\prime$ satisfying  $g_k^\prime=\gamma g_k\gamma^{-1}$. Then all marking
equivalent surfaces form a marking equivalence class $[S,\Gamma]$ representing
the Riemann surface $S$.

The set of all marking equivalence classes of the closed and compact Riemann surfaces
in genus $g$ forms the Teichm\"uller space ${\cal T}_g$. The real dimension of ${\cal T}_g$
like vector space equals to $6g-6$ in accordance with the Riemann--Roch theorem.
In our case, the Riemann surfaces result in the subset of total ${\cal T}_2$.

\section{Structure of Parameter Space ${\cal A}$}

\subsection{Symplectic Two-Form}
 
A hyperbolic Riemann surface of genus $g$ without boundary always contains a system of
$3g-3$ simple closed geodesics that are neither homotopic to each other nor homotopically
trivial. The cut along these geodesics always decomposes surface into $2g-2$ pairs of pants
(three-holed spheres), playing a role of natural building blocks for Riemann surface~\cite{Bu}.

In the case at hand, surface $S$ is two-holed torus which can be decomposed into two pairs
of pants by a system of three closed geodesics. This surgery results in computing the
Fenchel--Nielsen (FN) parameters: lengths $\ell_k$ of these geodesics and the corresponding
twists $\tau_k$ (see \cite{IT,Bu}).

For one of possible pants decompositions, we have found these quantities~\cite{Naz2}:
\begin{eqnarray}
\ell_{1,2}=2~{\rm arccosh}~{\frac{a^2}{1-a^2}},&\quad
&\ell_3=2\ln{\frac{1+a}{1-a}},\\
\tau_{1,2}={\rm arccosh}\left[\frac{2a^2-1}{a^2(1-b^2)}-1\right],&\quad
&\tau_3=\ln{\frac{1+a}{1-a}}.
\end{eqnarray}

Since the Teichm\"uller space ${\cal T}_2$ is homeomorphic to $\mathbb{R}^6$, one can
identify the FN variables with global coordinates on it. Moreover, the Teichm\"uller space
carries additional structure, namely, the  Weil--Petersson (WP) symplectic two-form. Due to
a theorem by Wolpert~\cite{IT,Wo08}, WP-form for compact closed Riemann surfaces of genus $g$
takes on a particularly simple form in terms of FN variables,
\begin{equation}\label{wwp}
\omega_{\rm WP}=\frac{1}{2}\sum\limits_{k=1}^{3g-3}\rmd\ell_k\wedge\rmd\tau_k,
\end{equation}
which is invariant with respect to any pants decomposition. Introducing $\theta_k=2\pi\tau_k/\ell_k$,
the simple Dehn twist $\theta_k\to\theta_k+2\pi$ gives us isometrically the same surface.

Substituting the expressions for $\ell_k$ and $\tau_k$ into (\ref{wwp}), WP-form becomes
\begin{equation}\label{wwp2}
\omega_{\rm WP}=\frac{8a}{(1-a^2)(2a^2\cos^2{\tilde\alpha}-1)}\rmd a\wedge \rmd\tilde\alpha.
\end{equation}

To verify the uniqueness of last formula, we can consider another pants decomposition.
In terms of our parameters, a new decomposition simply leads to replacement,
\begin{equation}\label{z2}
a\leftrightarrow b,\qquad \tilde\alpha\leftrightarrow-\tilde\alpha,
\end{equation}
in length and twist functions of previous decomposition. Although the set of new functions
is obtained, the resulting two-form $\omega_{\rm WP}$ remains the same.

Thus, domain ${\cal A}$ of admissible parameters $(a,\alpha)$ is symplectic manifold 
$({\cal A},\omega_{\rm WP})$. We may also treat two-form (\ref{wwp2}) as an area
element of ${\cal A}$.

\subsection{Isoperimetric Orbits}

Since the hyperbolic area of admissible octagons is always equal to $4\pi$, a simplest
way to control the surface changes globally consists in consideration of octagon perimeter:
\begin{equation}\label{Per}
P=8~{\rm arccosh}~\frac{1-a^2b^2+\sqrt{(1-a^2)^2+(1-b^2)^2}}{(1-a^2)(1-b^2)}.
\end{equation}

Although perimeter $P$ is obviously invariant under automorphism and pants decomposition, it can also
take on the same value for different pairs $(a,\alpha)$. We are aiming to describe the corresponding orbits.
On the contrary to our case, two parameters of  {\it flat} octagon with the same automorphism and
the homothety property are exactly defined by fixing its area and perimeter. 

For further calculations it is useful to introduce two auxiliary quantities:
\begin{equation}\label{varconv}
T\equiv\tanh{(P/16)}, \qquad \varepsilon=\pm1,
\end{equation}
where the latter one reflects an existence of two sheets in ${\cal A}$ labeled by
${\rm sign}~\tilde\alpha$.

For a given $T(P)$, maximal and minimal values of $a$ at $\tilde\alpha=0$ are
\begin{equation}\label{apm}
a_\pm(T)=\frac{1}{2}\sqrt{2+T^2\pm\sqrt{(2+T^2)^2-8}}.
\end{equation}

Solution to equation $(2+T^2)^2=8$ is $T_{\rm reg}=\sqrt{2\sqrt{2}-2}$ that results in
quantities $P_{\rm reg}=8~{\rm arccosh}~(5+4\sqrt{2})$ and $a_{\rm reg}=2^{-1/4}$
corresponding to the regular octagon. Thus, $P_{\rm reg}\approx24.457$ is a minimal value of $P$
among possible ones and trajectory in ${\cal A}$ for $P_{\rm reg}$ is contracted to a point.

Isoperimetric orbits can then be parametrized as
\begin{eqnarray}
a(T,\varphi)&=&\frac{1}{2}\sqrt{2+T^2+\cos{\varphi}\sqrt{(2+T^2)^2-8}},
\label{asol}\\
\tilde\alpha(T,\varphi)&=&\arctan{\frac{\sqrt{2}\sqrt{(2+T^2)^2-8}\sin{\varphi}}
{2\sqrt{3T^2-2-\cos{\varphi}\sqrt{(2+T^2)^2-8}}}},
\label{alf}
\end{eqnarray}
where cyclic variable $\varphi\in[0,2\pi)$ is used.

\begin{figure}
\begin{center}
\includegraphics[width=6.2cm]{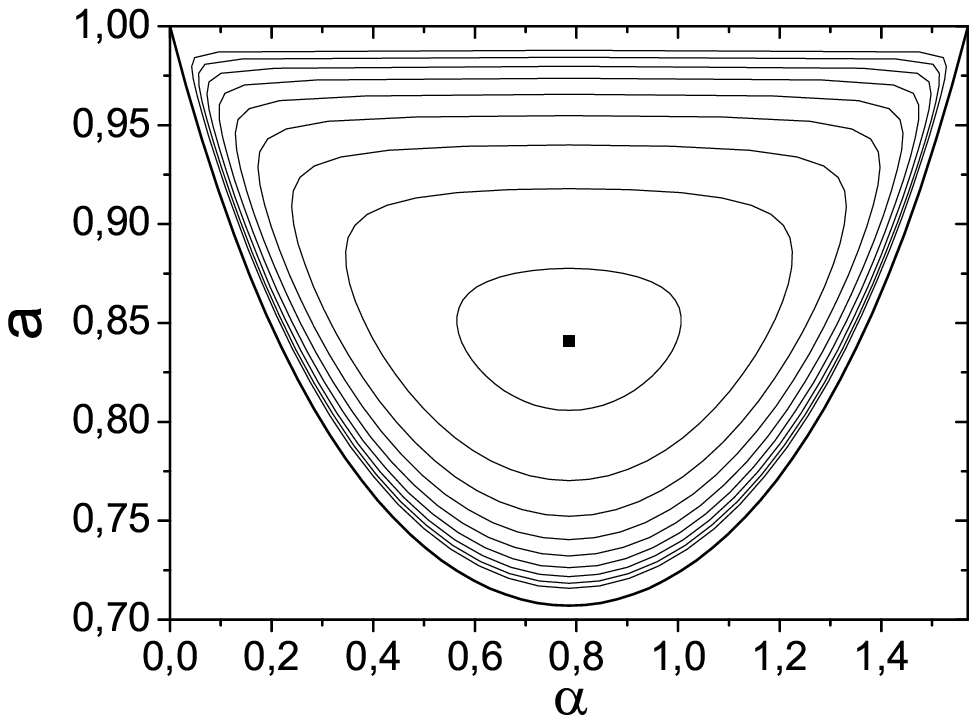}\ \
\includegraphics[width=5.9cm]{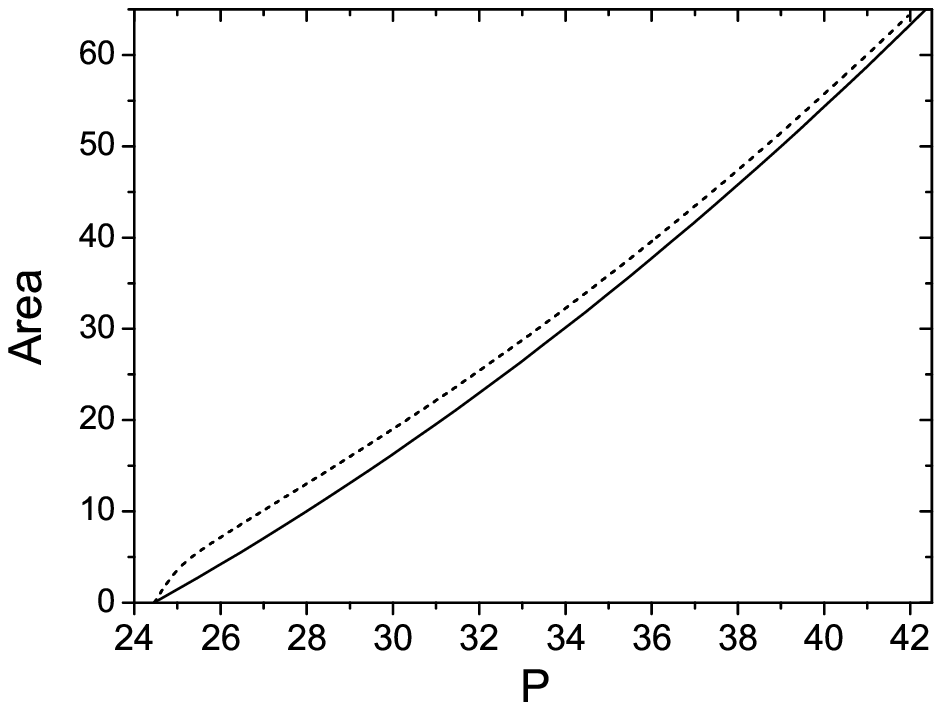}
\end{center}
\vspace*{-3mm}
\caption{\small Left panel: Parameter space ${\cal A}$ and orbits of constant perimeter.
Right panel: WP-Area of domain bounded by isoperimetric curve. Solid curve is computed
numerically; dashed curve is given by analytic approximation (\ref{Alog}).}
\end{figure}

These orbits are shown in Fig.~2 (left panel)  for $P$ from $P=25$ to $P=41$ with step 2. Although
a role of octagon perimeter is unclear from physical point of view, it is important that the
isoperimetric constraint guarantees the dense covering of ${\cal A}$, which has an unique shape, by
the corresponding orbits. This fact allows us to quantize $({\cal A},\omega_{\rm WP})$ in a spirit
of \cite{Hurt}. To realize it, we should identify the Weil-Petersson area $A_{\rm WP}(P)$ of domain
in ${\cal A}$, bounded by curve for fixed $P$, with an action variable, that is, integral of
``motion''. Then, quantization has to give us the number of quantum ``cells'' inside of the domain.
It seems enough if additional physics (except quantization) is not used. However, we shall require
SO(2,1) symmetry of dynamics in accordance with Lorentzian (2+1)-dimensional gravity.

Using the symplectic form (\ref{wwp2}), let us define WP-area:
\begin{eqnarray}
A_{\rm WP}(P)&\equiv&\int_{P={\rm const}}\omega_{\rm WP}\nonumber\\
&=&\int_{x_-(T)}^{x_+(T)}
\frac{8\rmd x}{(1-x)\sqrt{2x-1}}~{\rm arccosh}~f(x,T),
\label{AA}
\end{eqnarray}
where $x\equiv a^2$; functions $x_\pm(T)\equiv a^2_\pm(T)$ are determined by (\ref{apm});
\begin{equation}
f(x,T)=\frac{1}{\sqrt{1-T^2}}\sqrt{\frac{2x-1}{x}}\sqrt{\frac{T^2-x}{T^2-2x+1}},
\end{equation}
and convention (\ref{varconv}) is applied.

Analytic estimation of WP-area is made in Appendix~A. We show in Fig.~2 (right panel)
that $A^{(0)}_{\rm WP}(P)$ tends to numerically computed $A_{\rm WP}(P)$ at large $P$, but
behavior at $P\to P_{\rm reg}$ looks incorrect. However, we shall see that the derivative
$\rmd A_{\rm WP}(P)/\rmd P$ can be calculated explicitly.

\subsection{Global Canonical Variables}

Finding the canonical variables for isoperimetric orbits, let us define
\begin{equation}
Q(x,T)=\frac{\sqrt{2}}{4}\frac{1-T^2}{\sqrt{T^2-x_-}}\left[
\frac{F(u,k)}{1-T^2}-\frac{\Pi(u,\nu_1,k)}{1-x_-}+\frac{\Pi(u,\nu_2,k)}{T^2-2x_-+1}\right],
\end{equation}
where $F$ and $\Pi$ are elliptic integrals of the first and third kind, respectively.

Amplitude $u$, module $k$, and parameters $\nu_1$ and $\nu_2$ are given by
\begin{eqnarray}
&&u=\sqrt{\frac{x-x_-}{x_+-x_-}},\qquad
k=\sqrt{\frac{x_+-x_-}{T^2-x_-}},\nonumber\\
&&\nu_1=\frac{x_+-x_-}{1-x_-},\qquad
\nu_2=2\frac{x_+-x_-}{T^2-2x_-+1}.
\end{eqnarray}

Then, symplectic WP-form becomes
\begin{equation}\label{WW}
\omega_{\rm WP}=\varepsilon\rmd Q(x,T)\wedge\rmd P.
\end{equation}

Integrating over $x$, we see that
\begin{eqnarray}\label{dAWP}
\frac{\rmd A_{\rm WP}}{\rmd P}&\equiv&\oint_{P={\rm const}}\varepsilon\rmd Q(x,T)
=\int_{x_-}^{x_+}\rmd Q(x,T)-\int_{x_+}^{x_-}\rmd Q(x,T)\nonumber\\
&=&2Q(x_+,T),
\end{eqnarray}
where amplitude $u=1$ results in the complete elliptic integrals.

Defining an action variable (or ``angular momentum'') as
\begin{equation}\label{AM}
J(P)=\frac{1}{4\pi}A_{\rm WP}(P),
\end{equation}
we find the angle variable $\Phi$ from equation $\omega_{\rm WP}=2\rmd\Phi\wedge\rmd J$;
the Poisson bracket is then $\{J,\Phi\}_{\rm WP}=1$. One first has that
\begin{equation}
\rmd\Phi=\frac{\pi\varepsilon}{Q(x_+,T)}\rmd Q(x,T),\qquad
\oint_{P={\rm const}}\rmd\Phi=2\pi,
\end{equation}
where the rule of calculation of integrals containing $\varepsilon$ (see (\ref{dAWP})) is applied.

At fixed $P$, we come to expression for the angle variable:
\begin{equation}\label{Phi}
\Phi=\pi\varepsilon\frac{Q(x,T)}{Q(x_+,T)},\quad
\Phi\in[-\pi,\pi].
\end{equation}

Now it seems trivially to quantize the system in terms of $J$ and $\Phi$ what leads
immediately to estimation (in the Planck units) for relatively large $n$:
\begin{equation}
A_{\rm WP}\sim4\pi n,\qquad n\in\mathbb{N}.
\end{equation}
We specify this formula below due to consideration of the relativity theory.

\section{Relativistic Kinematics}

At this stage, geometrodynamics of Riemann surface within the considered model is
ambiguous because of purely gauge nature. There are only the geometric constraints
defining the ``physical sector'' of parameters variety and no definitions of time and
the Hamiltonian function having the physical meaning and generating an evolution of
$J$ and $\Phi$. However, we appeal here to (2+1)-dimensional gravity
where ${\rm SO}(2,1)\sim{\rm SU}(1,1)$ plays a role of the Lorentz group. It allows
us to construct a dynamical model with the same symmetry as follows.

Combining $J$ and $\Phi$, we extend the set of observables up to
\begin{equation}\label{su}
J_0=J,\quad
J_\pm=\sqrt{J^2-C}\exp{(\mp\rmi\Phi)},
\end{equation}
where $C$ is a constant such that $J^2\geqslant C\geqslant 0$.

The Poisson algebra of new variables is $\mathfrak{su}(1,1)$ Lie algebra:
\begin{equation}
\{J_+,J_-\}_{\rm WP}=2\rmi J_0,\qquad
\{J_\pm,J_0\}_{\rm WP}=\pm\rmi J_\pm.
\end{equation}

More generally, generators $J_{0,\pm}$ may be replaced by an infinite
number of quantities $L_n=J\exp{(\rmi n\Phi)}$, $n\in{\mathbb Z}$,
generating the Witt algebra.

Since the evolution of conservative system is usually described by canonical transformations,
we find that indeed SU(1,1) transformations are canonical transformations of a given system.
Let us introduce the matrix
\begin{equation}
{\cal M}=\left(
\begin{array}{cc}
J_0&J_+\\
J_-&J_0
\end{array}
\right),
\end{equation}
whose determinant $C\equiv J^2_0-J_+J_-$ is the Casimir of $\mathfrak{su}(1,1)$ algebra.

Action of matrix $U\in{\rm SU}(1,1)$ looks like ${\cal M}\mapsto\tilde{\cal M}=U{\cal M}U^\dag$
and preserves the Casimir, $\det\tilde{\cal M}=C$.

\begin{figure}
\begin{center}
\includegraphics[width=6.0cm]{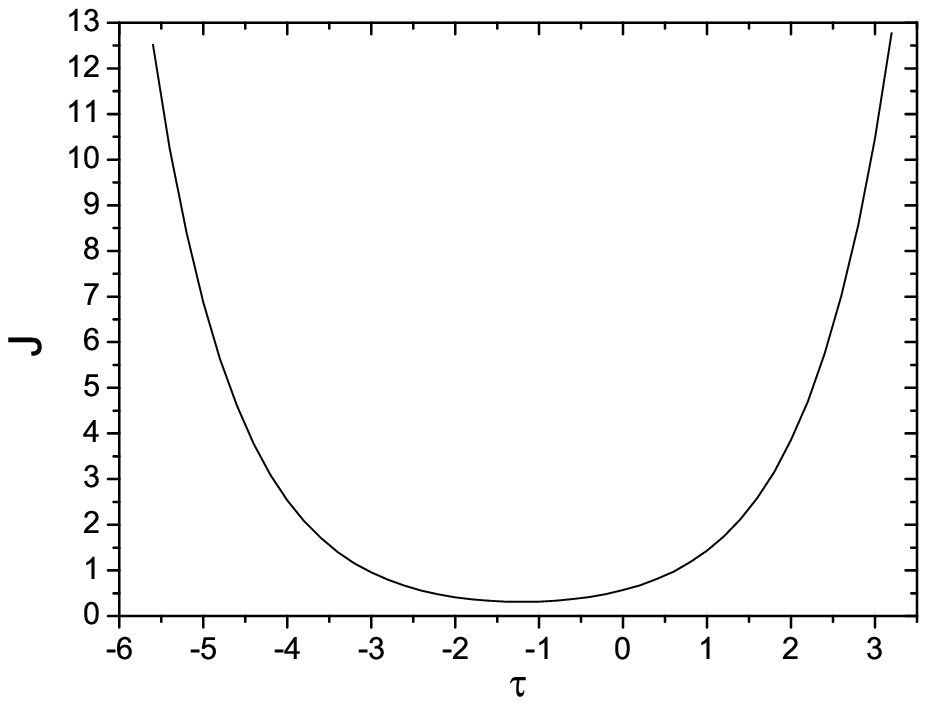}\ \
\includegraphics[width=6.3cm]{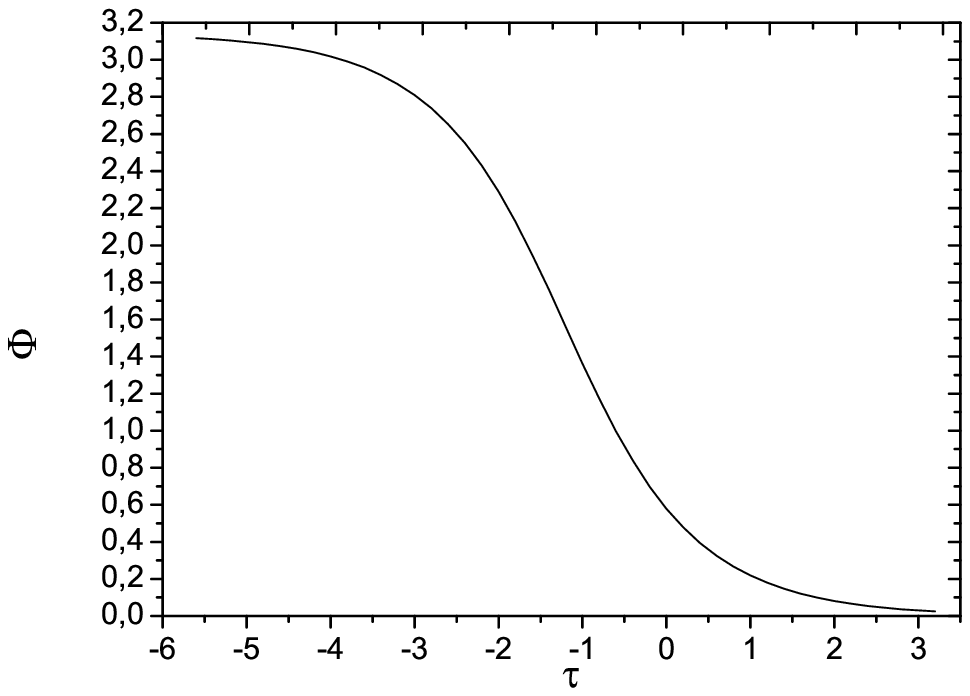}
\end{center}
\vspace*{-3mm}
\caption{\small Angular momentum $J$ and angle $\Phi$ as functions of
internal time $\tau$; $C=0$ and initial data are $a_0=0.8$, $\alpha_0=\pi/3$. The
surface starts with an infinite WP-area at $\tau\to-\infty$, contracts and bounces
to expand again to infinite area at $\tau\to\infty$.}
\end{figure}

Relativistic kinematics of the surface can be generated by the boost:
\begin{equation}
U_\tau=\left(\begin{array}{cc}
\cosh{(\tau/2)}&\sinh{(\tau/2)}\\
\sinh{(\tau/2)}&\cosh{(\tau/2)}
\end{array}\right).
\end{equation}
Computing ${\cal M}(\tau)=U_\tau{\cal M}(0)U^\dag_\tau$, one derives the trajectories
for $J_{0,\pm}$ or, equivalently, for $J$ and $\Phi$:
\begin{eqnarray}
J(\tau)&=&\breve{J}\cosh{\tau}+\sqrt{\breve{J}^2-C}\sinh{\tau}\cos{\breve{\Phi}},\label{Jt}\\
\Phi(\tau)&=&
\arccos{\frac{\breve{J}\sinh{\tau}+\sqrt{\breve{J}^2-C}\cosh{\tau}\cos{\breve{\Phi}}}{\sqrt{J^2(\tau)-C}}},
\label{Phit}
\end{eqnarray}
where $\breve{J}=J[P(a_0,\alpha_0)]$, $\breve{\Phi}=\Phi[a_0,P(a_0,\alpha_0)]$ are
the values of functions (\ref{AM}), (\ref{Phi}) at $\tau=0$.

On the other hand, the evolution can be described by the Hamiltonian equations:
\begin{eqnarray}
\partial_\tau J&\equiv&\{J,H\}_{\rm WP}=\sqrt{J^2-C}\cos{\Phi},\\
\partial_\tau\Phi&\equiv&\{\Phi,H\}_{\rm WP}=-\frac{J}{\sqrt{J^2-C}}\sin{\Phi},
\end{eqnarray}
where the Hamiltonian
\begin{equation}\label{H}
H(J,\Phi,C)=\sqrt{J^2-C}\sin{\Phi}
\end{equation}
belongs to $\mathfrak{so}(2,1)$ algebra.

Of course, the form of the Hamiltonian depends generally on the physical problem under
consideration. In relativistic cosmology, different scenarios lead to modifications
of (\ref{H}) which are discussed, for instance, in \cite{LM}.

In our case, time dependence of functions $J(\tau)$, $\Phi(\tau)$ is sketched in Fig.~3.
The figures demonstrate a ``bounce'' in parameter space. Finding the zeroes of derivatives
of (\ref{Jt}), (\ref{Phit}) with respect to $\tau$, the bounce characteristics are
\begin{eqnarray}
&&\tau_{\rm b}=-{\rm arctanh}\left(\frac{\sqrt{\breve{J}^2-C}}{\breve{J}}\cos{\breve{\Phi}}\right),
\nonumber\\
&&J_{\rm b}=\sqrt{\breve{J}^2\sin^2\breve{\Phi}+C\cos^2\breve{\Phi}},\quad
\Phi_{\rm b}=\varepsilon\frac{\pi}{2}.
\end{eqnarray}

\begin{figure}
\begin{center}
\includegraphics[width=8cm]{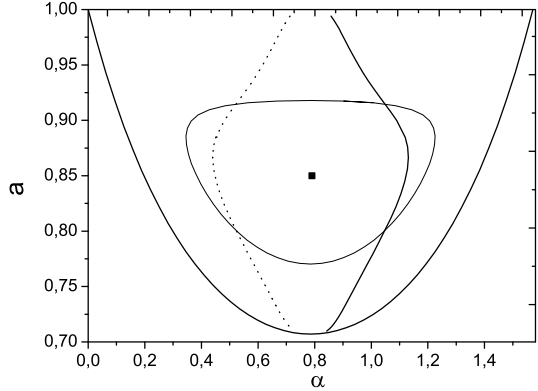}
\end{center}
\vspace*{-3mm}
\caption{\small The trajectory of Riemann surface evolution for $a_0=0.8$,
$\alpha_0=\pi/3$ and $C=0$. Bold line corresponds to $\varepsilon=1$, dashed
line is for $\varepsilon=-1$. Closed curve is the isoperimetric orbit determined
by initial data.}
\end{figure}

In principal, we are interested here in the form of trajectory in parameter space ${\cal A}$ without
specification of evolution parameter or time. It can be found by means of constraint
\begin{equation}\label{Hc}
H\{J[P(a,\alpha)],\Phi[a,P(a,\alpha)],C\}=E,\qquad
(a,\alpha)\in{\cal A},
\end{equation}
where constant $E$ is the value of Hamiltonian function $H$ for initial data.

Although Eq.~(\ref{Hc}) says at first sight that the points of trajectory are simply determined from
an abstract equation $h(a,\alpha,C,E)=0$ with additional constants $C$ and $E$, we would like to
emphasize its structure reflecting the chain of our buildings.

In terms of $a$ and $\alpha$ the trajectory looks like in Fig.~4, where influence of $C$ is neglected
because of quantum nature assumed. It is interesting to note that the system geometry does not tend to
the regular octagon having a maximal information entropy. At the values $a_{\rm min}=1/\sqrt{2}$ and
$a_{\rm max}=1$ corresponding to the infinite past and the infinite future time, angle $\alpha$ of
hyperbolic octagon reaches the same value $\pi/4$. Thus, geometry with $\alpha=\pi/4$ is exceptional in
our model. We would like also to note that there are two configurations for $\varepsilon=\pm1$ (left
and right with respect to $\tilde\alpha=0$), which are not mixed during whole evolution generated by
the pure boost. The chosen preference is preserved from the origin to the end. However, we have already
seen that there is the diffeomorphism generated by conservation condition of hyperbolic octagon perimeter,
which allows us the transition between ``phases'' with $\varepsilon=\pm1$. Combining it with an action of
the boost, it is possible to construct a novel scenario of classical geometrodynamics, even in the case
when the Riemannian metric of the parameter (or moduli) space is undetermined. 

Coming back to quantization problem, the generator $J_0$ and its spectrum describes WP-area.
For this reason its eigenvalues should be discrete and positive. We choose the irreducible representation
(\ref{su11}) with standard basis diagonalizing the Casimir and $J_0$ and with minimal positive spin
$j=1/2$. It leads straightforwardly to the spectrum (in the Planck units):
\begin{equation}
A_{\rm WP}=4\pi\left(n+\frac{1}{2}\right),\qquad
n\in\mathbb{N}.
\end{equation}
In classical picture, $A_{\rm WP}=0$ for $P_{\rm reg}$, while $A_{\rm WP}$ is always non-vanishing at
quantum level for the system with a given topology. It means that the regular octagonal configuration
is not achieved because of quantum effect.

Although the basis of $\mathfrak{su}(1,1)$ is enough to quantize $A_{\rm WP}$ as a global characteristic of
${\cal A}$, it looks insufficient to apply for finding the spectra of other geometric observables. Further
investigations are still needed.

\section{Conclusions}

Here, we pay great attention to the structure of admissible parameters space ${\cal A}$
determining the geometry of the Riemann surface in genus two with an order four automorphism.
First, using the Weil--Petersson (WP) geometry, two-dimensional space ${\cal A}$ is equipped
with the fundamental symplectic two-form. Further, we perform the dense covering of ${\cal A}$
by the orbits generated by the isoperimetric constraint which is imposed on fundamental domain
of the surface. It is argued that an existence of these orbits is due to the Riemann surface
definition. The canonically conjugate action--angle variables for isoperimetric orbits are
found by identifying WP-area of domain bounded by the orbit with the action variable.
As the result, we take on a possibility to construct relativistic model in terms of special
invariants without knowing the Riemannian metric of ${\cal A}$.

To build the physically meaningful model, we extend the set of global canonical variables up
to generators of $\mathfrak{su}(1,1)$ algebra. This trick leads to appearance of the
Casimir playing a role of additional parameter. We can only assume that its value should be
minimal and non-zero. However, it leads after quantization to non-vanishing discrete spectrum
of WP-area in a contrast with initial theory where WP-area becomes zero for the Riemann
surface associated with the regular hyperbolic octagon. In any case, we should remember that
WP-area determines the surface geometry up to diffeomorphism in classical theory.
We may need to use the infinite Witt and Virasoro algebras instead of $\mathfrak{su}(1,1)$
in order to describe the system spectrum.

We also consider relativistic kinematics or free geometrodynamics of the Riemann surface,
generated by the Lorentz boost which acts on the constructed generators of $\mathfrak{su}(1,1)$
algebra. The time dependence of global variables leads to ``big bounce'' scenario and is
similar for quantities of different origin. However, solving equations with respect to
the surface parameters, we have obtained the trajectory (independent on time definition) in
space ${\cal A}$. In this picture, the system does not tend to reach the regular octagon
configuration corresponding to the maximal information entropy~\cite{Naz} and preserves some
kind of $\mathbb{Z}_2$ symmetry related to admissibility range of angle variable during whole
evolution.

\section*{Acknowledgments}

A.N. is deeply indebted to A.M.~Gavrilik (BITP, Kiev) and I.V.~Mykytiuk (IAPMM, Lviv)
for fruitful discussions of mathematical aspects of the problem. 

\appendix

\section{Weil--Petersson Area Estimation}

To evaluate the integral (\ref{AA}) analytically, we use the asymptotic expansion:
\begin{equation}
{\rm arccosh}~z=\ln{(2z)}-\frac{1}{2\cdot2z^2}-\frac{1\cdot3}{2\cdot4\cdot4z^4}
-\frac{1\cdot3\cdot5}{2\cdot4\cdot6\cdot6z^4}-\ldots.
\end{equation} 

Limiting ourselves by accounting for logarithmic term only, we obtain
\begin{equation}\label{Alog}
A^{(0)}_{\rm WP}(P)=F(T,\sqrt{2x_+(T)-1})-F(T,\sqrt{2x_-(T)-1}),
\end{equation}
where 
\begin{eqnarray}\hspace*{-5mm}
F(T,\xi)&=&16\ln{\frac{2}{\sqrt{1-T^2}}}~{\rm arctanh}~\xi+8\ln{\xi}\ln{(1+\xi)}
+8~{\rm dilog}~\xi\nonumber \\
&&+8~{\rm dilog}~(1+\xi)
+4\sum\limits_{\epsilon=\pm1}\left\{~{\rm dilog}~\frac{T+\epsilon\xi}{T-\epsilon}
-{\rm dilog}~\frac{T+\epsilon\xi}{T+\epsilon}
\right.
\nonumber\\
&&+
{\rm dilog}~\left[\frac{1-\xi}{2}+\epsilon\rmi\frac{1+\xi}{2}\right]
-{\rm dilog}~\left[\frac{1+\xi}{2}+\epsilon\rmi\frac{1-\xi}{2}\right]
\nonumber\\
&&\left.+{\rm dilog}~\frac{\sqrt{2T^2-1}+\epsilon\xi}{\sqrt{2T^2-1}+\epsilon}
-{\rm dilog}~\frac{\sqrt{2T^2-1}+\epsilon\xi}{\sqrt{2T^2-1}-\epsilon}
\right\}.
\end{eqnarray}

Here, the dilogarithm function is defined by the following series:
\begin{equation}
{\rm dilog}~z=\sum\limits_{p=1}^\infty\frac{(1-z)^p}{p^2}.
\end{equation}

\section{Basis of SU(1,1)}

We use the usual basis of SU(1,1) diagonalizing both the Casimir and $J_0$.
The action of $\mathfrak{su}(1,1)$ generators on this orthonormal basis is
\begin{eqnarray}
&C|j,m\rangle=j(j-1)|j,m\rangle,\qquad
J_0|j,m\rangle=m|j,m\rangle,&\label{su11}\\
&J_+|j,m\rangle=\sqrt{(m-j+1)(m+j)}|j,m+1\rangle,&\nonumber\\
&J_-|j,m\rangle=\sqrt{(m-j)(m+j-1)}|j,m-1\rangle.&\nonumber
\end{eqnarray}
There are two types of (discrete) representations: the positive series with $1/2\leq j\leq m=j+\mathbb{N}$;
and the negative one with $-1/2\geq j\geq m=j-\mathbb{N}$. Here we restrict our consideration by
the irreducible representation of positive spin $j$.


\end{document}